\documentclass[preprint,12pt]{elsarticle}

\usepackage{amssymb}

\usepackage{color}

\journal{}

\begin{document}

\begin{frontmatter}



\title{A generalized public goods game with coupling of individual ability and project benefit}


\author{Li-Xin Zhong$^{a}$}\ead{zlxxwj@163.com}
\author {Wen-Juan Xu$^b$}
\author {Yun-Xin He $^{a}$}
\author{Chen-Yang Zhong$^{c}$}
\author {Rong-Da Chen $^{a}$}
\author {Tian Qiu$^d$}
\author {Yong-Dong Shi$^e$}
\author {Fei Ren$^f$}

\address[label1]{School of Finance and Coordinated Innovation Center of Wealth Management and Quantitative Investment, Zhejiang University of Finance and Economics, Hangzhou, 310018, China}
\address[label2]{School of Law, Zhejiang University of Finance and Economics, Hangzhou, 310018, China}
\address[label3]{Yuanpei College, Peking University, Beijing, 100871, China}
\address[label4]{School of Information Engineering, Nanchang Hangkong University, Nanchang, 330063, China}
\address[label5]{Research Center of Applied Finance, Dongbei University of Finance and Economics, Dalian, 116025, China}
\address[label6]{School of Business and Research Center for Econophysics, East China University of Science and Technology, Shanghai, 200237, China}

\begin{abstract}
Facing a heavy task, any single person can only make a limited contribution and team cooperation is needed. As one enjoys the benefit of the public goods, the potential benefits of the project are not always maximized and may be partly wasted. By incorporating individual ability and project benefit into the original public goods game, we study the coupling effect of the four parameters, the upper limit of individual contribution, the upper limit of individual benefit, the needed project cost and the upper limit of project benefit on the evolution of cooperation. Coevolving with the individual-level group size preferences, an increase in the upper limit of individual benefit promotes cooperation while an increase in the upper limit of individual contribution inhibits cooperation. The coupling of the upper limit of individual contribution and the needed project cost determines the critical point of the upper limit of project benefit, where the equilibrium frequency of cooperators reaches its highest level. Above the critical point, an increase in the upper limit of project benefit inhibits cooperation. The evolution of cooperation is closely related to the preferred group-size distribution. A functional relation between the frequency of cooperators and the dominant group size is found.
\end{abstract}


\end{frontmatter}


\section{Introduction}
\label{sec:introduction}
The occurrence and maintenance of cooperation in a competitive setting have drawn much attention of biologists, economists, sociologists, mathematicians and statistical physicists\cite{perc10,hadzibeganovic1,hadzibeganovic2,qiu1,hadzibeganovic3,hadzibeganovic4,hadzibeganovic30}, which is similar to the diffusion systems and the complex systems in the physical world\cite{johnson10,allegrini10,aquino10,geneston10,courbage10,courbage11,han30}. In order to find the internal mechanism of the Nash equilibrium and the flourish of cooperation among completely rational individuals, evolutionary game theory and quite a few classical game models, such as the prisoner's dilemma (PD) and the snowdrift game (SG), have been employed to model the evolution of altruistic behaviour\cite{szabo10,xu10,nowak10,hauert10,nax1,szabo1,schweitzer1,schweitzer2,szolnoki4}. The PD is a standard metaphor to explain the evolution of cooperation through pairwise interactions. For group interactions, the public goods game (PGG) represents a straightforward generalization of the PD. The SG is also a cooperative game model describing pairwise interactions. The only difference between the PD and the SG is the payoff matrix. In the PD, a cooperator gets nothing. In the SG, a cooperator has a net gain after deducting the cost of cooperation. The N-person snowdrift game (NSG) represents a straightforward generalization of the SG.

In the original PGG\cite{szabo2,helbing1}, each individual has to decide whether to make a contribution to a common pool or not. As all the individuals have made their decisions, the total investment in the common pool is multiplied and distributed equally among all the members in the interacting group. Those who contribute to the common pool are cooperators and those who make no contribution are defectors. A rational analysis will result in such a bad scenario where nearly all the individuals contribute nothing to the common pool in order to obtain a higher personal gain and the tragedy of the commons occurs. In the PGG, each cooperator's contribution is the same and predefined. An increase in the number of cooperators in the interacting group does not change each cooperator's contribution but leads to a rise of the total cost and the total benefit. Different from the payoff functions in the PGG, in the NSG\cite{zheng1,souza20}, having finished a task, each member of the interacting group gets the same and predefined benefit while the predefined cost is evenly shared by cooperators. Therefore, in the NSG, the total benefit increases while the total cost does not change with the rise of the group members.

However, in real society, the evolutionary mechanism in the PGG and that in the NSG may coexit\cite{schweitzer3}. The upper limit of individual ability and the upper limit of project benefit may result in such a scenario: facing a heavy task, an individual can do nothing because of his limited ability. Only when there are quite a few cooperators in the interacting group can the task be finished and the project benefit be obtained. For example, the task of moving a big and heavy rock or hunting a big creature can not be finished until there are enough cooperators in the interacting group. The impact of the critical mass or the start-up cost on the evolution of cooperation has been discussed by A. Szolnoki et al\cite{szolnoki1,szolnoki2}. In addition to that, the profit from finishing the heavy task depends on the potential benefits of the project and the characteristics of the individuals who have the right to enjoy such benefits. The existence of the upper limit of individual benefit may result in such a scenario where the project benefits are partly wasted. For example, as a bridge has been built, its bearing capacity may not be fully exploited until there are quite a few individuals to-and-fro. As a group of wolves have captured a big deer, they usually enjoy a good meal and give up the leftovers. The profit from hunting a big creature can not be maximized until there are quite a few wolves in the group. As the project benefit has been maximized, an increase in the team members will lead to a decrease in the individual benefit. Although the start-up cost and the coupling of the PD game and the SG game have been discussed in refs.\cite{szolnoki1,szolnoki2,wang10,wang11}, the coupling effect of individual ability and project benefit on the evolution of cooperation and the coupling model of PGG and NSG are left unconsidered.

The coupling effect of different kinds of games on the evolution of cooperation is usually studied depending upon the threshold game models\cite{szolnoki2, perc30, szolnoki30, chen30, zhang30}. A. Szolnoki1 et al. have incorporated the start-up costs into the public goods game\cite{szolnoki2}. They have found that the existence of a threshold acting as an initial contribution to the common pool can promote the levels of cooperation effectively. M. Perc has incorporated the success-driven mechanism into the public goods game\cite{perc30}. He has found that the reproductive success of individuals promotes cooperation effectively irrespective of the interaction structure. A. Szolnoki1 et al. have introduced a level of payoff acting as a threshold for an individual to organize the public goods game\cite{szolnoki30}. They have found that such a mechanism can keep cooperation at a somewhat high level. X. J. Chen et al. have incorporated maximal endowments into the public goods game\cite{chen30}. They have found that an excessive abundance of common resources is detrimental to cooperation. J.L. Zhang et al. have incorporated an insurance covering the potential loss into the threshold public goods game\cite{zhang30}. They have found that an increase in the compensation from the insurance leads to more contributions. The role of other threshold parameters in the evolution of cooperation has been discussed in refs.\cite{cadsby30, dragicevic20, bach20, cadsby20, cadsby21, wang23, hsu20, croson20, marks20, mikkelsen20, Chen20, lu20}.

The evolution of individual strategies may not be sufficient for the occurrence and maintenance of cooperation among selfish individuals. The coevolutionary mechanism, such as the coevolution of the individual strategies and the interaction structures, may be seen as an effective way to promote cooperation. M. Perc et al. have reviewed the coevolutionary rules affecting the evolution of cooperation\cite{perc40}. The rules of mutual interaction, population growth, reproduction, mobility, reputation  and aging have a powerful effect on the evolution of cooperation.

To mimic the limitedness of individual ability and the potential benefits of the project, in the present we introduce four parameters, the upper limit of individual contribution, the upper limit of individual benefit, the needed project cost and the upper limit of project benefit into the original PGG. In addition to that, the evolution of the individual-level group-size preferences is also considered. Accompanied by the coevolution of the preferred group-sizes and the strategies of cooperation and defection, the coupling effect of individual ability and project benefit on the evolution of cooperation is extensively studied. We have three main findings.

(1) A higher level of individual contribution leads to a lower level of cooperation while a higher level of the upper limit of individual benefit leads to a higher level of cooperation. In the process of carrying out a heavy task with a given cost, if an individual's maximum contribution is limited, there should be more cooperators in finishing the task. The existence of the start-up cost promotes cooperation. As the upper limit of individual benefit increases, an individual is more possible to obtain a higher benefit, which leads to a higher level of cooperation.

(2) A higher level of the upper limit of project benefit leads to a lower level of cooperation. There exists a critical point of the upper limit of project benefit, below which the equilibrium frequency of cooperators changes little with the rise of the upper limit of project benefit and above which an increase in the upper limit of project benefit leads to a decrease in cooperation. The critical point is determined by the coupling of the upper limit of individual contribution and the needed project cost.

(3) A higher level of cooperation is in accordance with a smaller value of the dominant group size. The frequency of cooperators coevolves with the individual-level group size preferences. The occurrence of a higher level of cooperation is accompanied by the occurrence of a smaller dominant group size. A functional relation between the equilibrium frequency of cooperators and the dominant group size is found.

The generalized public goods game (GPGG) is presented in section 2. Simulation results and discussions are given in section 3. In section 4 we give a theoretical analysis on the relationship between the equilibrium frequency of cooperators and the dominant group size. In section 5 we summarize our conclusions.

\section{The model}
\label{sec:model}
The generalized public goods game (GPGG) is defined as follows. Assuming there is a project with the needed project cost $C_{pro}$ and the upper limit of project benefit $B^{max}_{pro}$. If there are enough cooperators in the interacting group, $n_C\ge n^{min}_C$, the project can be finished and the project benefit $B_{pro}$ depends on the upper limit of project benefit $B^{max}_{pro}$, the upper limit of individual benefit $B^{max}_I$ and the number of individuals $n$ in the interacting group. On condition that $nB^{max}_I\le B^{max}_{pro}$, $B_{pro}=nB^{max}_I$. On condition that $nB^{max}_I>B^{max}_{pro}$, $B_{pro}=B^{max}_{pro}$. If there are not enough cooperators in the interacting group, $n_C<n^{min}_C$, the project can not be finished and the project benefit is $B_{pro}=0$.

The threshold of the number of cooperators $n^{min}_C$ is determined by the needed project cost $C_{pro}$ and the upper limit of individual contribution $C^{max}_I$, which is satisfied with the equation $n^{min}_C=[\frac{C_{pro}}{C^{max}_I}]+1$ for $\frac{C_{pro}}{C^{max}_I}>[\frac{C_{pro}}{C^{max}_I}]$ and $n^{min}_C=[\frac{C_{pro}}{C^{max}_I}]$ for $\frac{C_{pro}}{C^{max}_I}=[\frac{C_{pro}}{C^{max}_I}]$. For $n_C\ge n^{min}_C$, a cooperator's contribution to the project is $C_C=\frac{C_{pro}}{n_C}$ and a defector's contribution to the project is $C_D=0$. For $n_C< n^{min}_C$, neither cooperators nor defectors make a contribution to the project, $C_C=C_D=0$.

The value of $n^{min}_C$ is closely related to the values of $C_{pro}$ and $C^{max}_I$. The relationship between $n^{min}_C$ and $\frac {C_{pro}}{C^{max}_I}$ is a step function. For  example, within the range of $C_{pro}\le C^{max}_I$, $n^{min}_C=1$, within the range of $1<\frac{C_{pro} }{C^{max}_I}\le 2$, $n^{min}_c=2$. Such an assumption is in accordance with the scenario where we have to finish a heavy and indivisible task. For example, if the road was blocked by a heavy rock, it is difficult for an individual to remove the roadblock alone. There should be more cooperative individuals to lift the heavy rock together. In order to catch a giant prey, such as an adult buffalo, a tiger needs more cooperative partners. Or else, such a heavy and indivisible task can not be finished successfully. Although the present model is too simplified to account for the real situations completely, it may become suitable for further researches in the direction of discontinuous task problems.

Each individual has a preference for the group-size $n$\cite{shi1,szolnoki3,zhang1,xu1}. At each time step, firstly, we randomly choose an individual $i$ with the preferred group-size $n_i$ from the total $N$ individuals. Then, from the left $N-1$ individuals we randomly choose another $n_i-1$ individuals. The payoff of individual $i$ is obtained, $P_i=P_C$ for a cooperator and $P_i=P_D$ for a defector. Thirdly, we randomly choose another individual $j$ and obtain his payoff $P_j$. With probability

\begin{equation}
\label{eq.1}
\omega_{i\gets j}=\frac{1}{1+e^{(P_i-P_j+\tau)/\kappa}},
\end{equation}
in which $\tau=\kappa=0.1$, individual i adopts individual j's strategy. Or else, individual i doesn't update his strategy. Fourthly, individual i compares his payoff $P_i$ with the averaged payoff of all the individuals $\bar P$. If $P_i<\bar P$, individual i randomly chooses another group-size $n'_i$ from the range of $n'_i\in [n_i-1,n_i+1]$ and updates his preferred group-size $n_i$ with group-size $n'_i$. If $P_i\ge\bar P$, individual i doesn't update his preferred group-size. Both the strategies and the preferred group-sizes are updated asynchronously.

The payoffs of cooperators and defectors are as follows. For $n_CC^{max}_I\ge C_{pro}$ and $B^{max}_{pro}\ge nB^{max}_I$,

\begin{equation}
\label{eq.1}
P_C=B^{max}_I-\frac{C_{pro}}{n_C}, P_D=B^{max}_I.
\end{equation}
For $n_CC^{max}_I\ge C_{pro}$ and $B^{max}_{pro}<nB^{max}_I$,

\begin{equation}
\label{eq.1}
P_C=\frac{B^{max}_{pro}}{n}-\frac{C_{pro}}{n_C}, P_D=\frac{B^{max}_{pro}}{n}.
\end{equation}
For $n_CC_I^{max}< C_{pro}^{max}$,
\begin{equation}
\label{eq.1}
P_C=P_D=0.
\end{equation}

The proposed GPGG model is a coupling model of the PGG and the NSG. Consider the case where all the individuals have the same preferred group-size $n$ and it does not range with time. Comparing the payoffs in the present model with that in the NSG, we find that the original NSG model can be obtained as a special case of the present model. In the NSG, the payoffs of a cooperator and a defector are satisfied with the equations $P_C=B_I-\frac{C_{pro}}{n_C}$ and $P_D=B_I$. In the present model, within the range of $n_CC^{max}_I\ge C_{pro}$ and $nB^{max}_I\le B^{max}_{pro}$, the payoffs of a cooperator and a defector are satisfied with the equations $P_C=B^{max}_I- \frac{C_{pro}}{n_C}$ and $P_D=B^{max}_I$. On condition that $C^{max}_I\sim\infty$ and $B_{pro}=nB_I$, the payoffs in the NSG can be obtained from the payoffs in the present model. Comparing the payoffs in the traditional PGG with that in the present model, we also find that the traditional PGG can be obtained as a special case of the present model. In the traditional PGG, the payoffs of a cooperator and a defector are satisfied with the equations $P_C=\frac{rn_Cx}{n}-x$ and $P_D=\frac{rn_Cx}{n}$, in which $r$ is the multiplication factor and $x$ is the contribution of a cooperator. In the present model, within the range of $n_CC^{max}_I\ge C_{pro}$ and $nB^{max}_I\ge B^{max}_{pro}$, the payoffs of a cooperator and a defector are satisfied with the equations $P_C=\frac{B^{max}_{pro}}{n}- \frac{C_{pro}}{n_C}$ and $P_D=\frac{B^{max}_{pro}}{n}$. On condition that $B^{max}_I\sim\infty$, $C_I=x$, $C_{pro}=n_C C_I$  and $B_{pro}= r C_{pro}$, the payoffs in the traditional PGG can be obtained from the payoffs in the present model.

Consider the case where the preferred group-sizes deviate from the $\delta$-function, due to the coexistence of n and n', which are satisfied with the inequalities $nB^{max}_I\ge B^{max}_{pro}$ and $n'B^{max}_I<B^{max}_{pro}$ respectively, the proposed model become the coupling model of the original PGG and the NSG. The payoff functions in the original PGG accompanied with the payoff functions in the NSG determine the evolutionary dynamics in the present model. The coupling model of the PD and the SG is also presented in ref.\cite{ma20,wang1}.

\section{Simulation results and discussions}
\begin{figure}
\includegraphics[width=8cm]{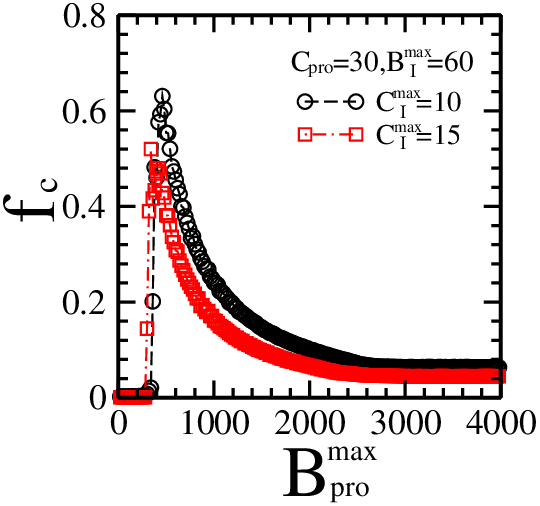}
\caption{\label{fig:epsart} the frequency of cooperators $f_C$ as a function of the upper limit of project benefit $B^{max}_{pro}$ for the upper limit of individual contribution $C^{max}_{I}$=10 (circles), 15 (squares). Other parameters are: the upper limit of individual benefit $B^{max}_I$=60, the needed project cost $C_{pro}$=30, the minimum group size $n_{min}=2$, the maximum group size $n_{max}=50$, the total number of individuals $N$=1000. The data are obtained by averaging over 10 runs and $10^3$ time steps after $10^4$ relaxation time in each run.}
\end{figure}

Starting with the basic setup where the strategies and the preferred group-sizes are randomly chosen. To explore the coupling effect of the upper limit of individual contribution $C^{max}_{I}$ and the upper limit of project benefit $B^{max}_{pro}$ on the prevalence of cooperation, in fig. 1 we plot the frequency of cooperators $f_c$ as a function of $B^{max}_{pro}$ for different $C^{max}_{I}$. For a given $C^{max}_{I}$=10, the critical point of $B^{maxc}_{pro}\sim 300$ is observed. As $B^{max}_{pro}$ increases from $B^{max}_{pro}\sim 0 $ to $B^{max}_{pro}\sim 300$, $f_c$ keeps its minimum value of $f_c\sim 0$. As $B^{max}_{pro}$ increases from $B^{max}_{pro}\sim 300 $ to $B^{max}_{pro}\sim 480$, $f_c$ has a sharp rise from $f_c\sim 0$ to $f_c\sim 0.64$. The prevalence of cooperation, i.e. $f_c\sim 0.64$, is observed at the point of $B^{maxc}_{pro}\sim 480$. As $B^{max}_{pro}$ increases from $B^{max}_{pro}\sim 480$ to $B^{max}_{pro}\sim 2000$, $f_c$ decreases continuously from $f_c\sim 0.64$ to $f_c\sim 0.1$. As $B^{max}_{pro}$ increases from $B^{max}_{pro}\sim 2000$ to infinity, $f_c$ has little change with the rise of $B^{max}_{pro}$. For a larger $C^{max}_{I}$, the critical point of $B^{maxc}_{pro}$ becomes smaller and the changing tendency of $f_c$ vs $B^{max}_{pro}$ changes little.

\begin{figure}
\includegraphics[width=10cm]{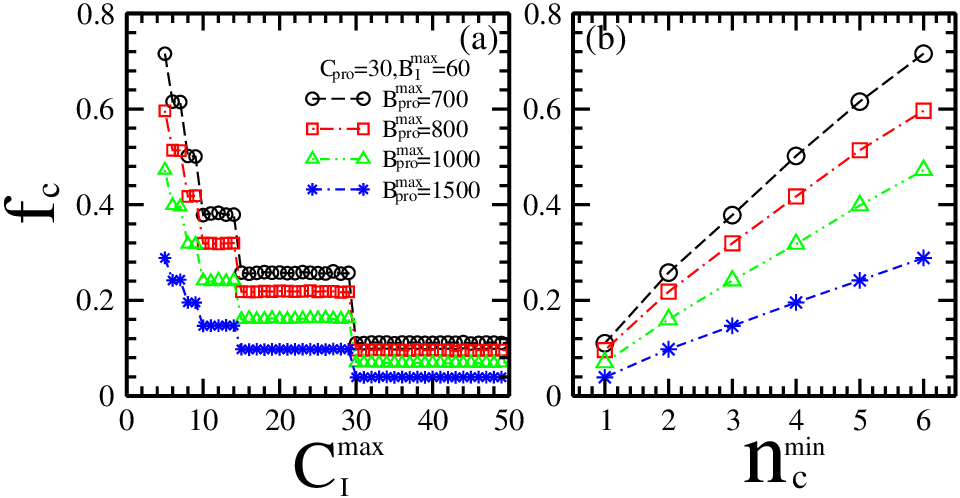}
\caption{\label{fig:epsart} the frequency of cooperators $f_c$ (a) as a function of the upper limit of individual contribution $C^{max}_{I}$ for the upper limit of project benefit $B^{max}_{pro}$=700 (circles), 800 (squares), 1000 (triangles), 1500 (stars); (b) as a function of the minimum number of cooperators $n^{min}_C$ for $B^{max}_{pro}$=700 (circles), 800 (squares), 1000 (triangles), 1500 (stars). Other parameters are: the upper limit of individual benefit $B^{max}_I$=60, the needed project cost $C_{pro}$=30, the minimum group size $n_{min}=2$, the maximum group size $n_{max}=50$, the total number of individuals $N$=1000. The data are obtained by averaging over 10 runs and $10^3$ time steps after $10^4$ relaxation time in each run.}
\end{figure}

The frequency of cooperator $f_c$ as a function of $C^{max}_{I}$ for different $B^{max}_{pro}$ is presented in fig. 2 (a). The dependence of $f_c$ on $C^{max}_{I}$ is like a step function. For a given $B^{max}_{pro}$=700, at the point of $C^{max}_{I}\sim 5$, $f_c\sim 0.72$. Between $C^{max}_{I}\sim 6$ and $C^{max}_{I}\sim 7$, $f_c\sim 0.61$. Between $C^{max}_{I}\sim 8$ and $C^{max}_{I}\sim 9$, $f_c\sim 0.5$. Between $C^{max}_{I}\sim 10$ and $C^{max}_{I}\sim 14$, $f_c\sim 0.38$. Between $C^{max}_{I}\sim 15$ and $C^{max}_{I}\sim 29$, $f_c\sim 0.25$. Within the range of $C^{max}_{I}\ge 30$, $f_c\sim 0.11$. For a larger $B^{max}_{pro}$, the changing tendency of the steps changes little while $f_c$ decreases with the rise $B^{max}_{pro}$.

The relationship between the cooperation level and the threshold  $n^{min}_C$ has been plotted in fig.2 (b). For a given $B^{max}_{pro}$ and $B^{max}_I$, $f_C$ increases with the rise of $n^{min}_C$. Within the whole range of $n^{min}_C\ge 1$, an increase in $\frac{B_{pro}}{B_I}$ leads to an overall decrease in $f_C$. How the parameter $n^{min}_C$ influences the cooperation level can be understood as follows\cite{szolnoki2}. In the interacting group, only when the number of cooperators is equal to or larger than $n^{min}_C$, the individuals in the group can get a net benefit, which leads to the local density of cooperators. As the interacting group size evolves to $n\sim n^{min}_C$, the individuals in the group can get a net benefit only when all the individuals are cooperators. Therefore, accompanied by the coevolutionary mechanism, the existence of the threshold parameter $n^{min}_C$ can promote the level of cooperation in the present model.

The results in fig. 1 and fig. 2 indicate that an optimal level of cooperation is determined by the coupling of the upper limit of individual contribution and the upper limit of project benefit. If an individual's maximum contribution to a heavy task is constrained, the level of cooperation is promoted. If the project benefit can be enjoyed by more people, the level of cooperation is inhibited.

\label{sec:results}
\begin{figure}
\includegraphics[width=8cm]{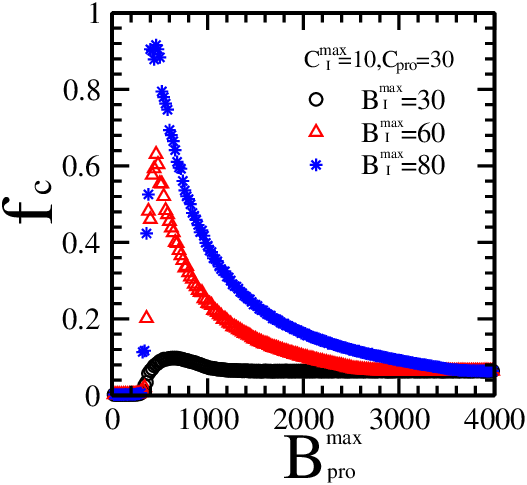}
\caption{\label{fig:epsart} the frequency of cooperators $f_C$ as a function of the upper limit of project benefit $B^{max}_{pro}$ for the upper limit of individual benefit $B^{max}_{I}$=30(circles), 60(squares), 80(triangles). Other parameters are: the needed project cost $C_{pro}$=30, the upper limit of individual contribution $C^{max}_{I}$=10, the minimum group size $n_{min}=2$, the maximum group size $n_{max}=50$, the total number of individuals $N$=1000. The data are obtained by averaging over 10 runs and $10^3$ time steps after $10^4$ relaxation time in each run.}
\end{figure}

As to the individual ability, the benefits and the costs are closely correlated with each other. In fig. 3, we plot the frequency of cooperators $f_c$ as a function of the upper limit of project benefit $B^{max}_{pro}$ for different values of the upper limit of individual benefit $B^{max}_I$. For a comparatively small $B^{max}_I$, i.e. $B^{max}_I=30$, the frequency of cooperators keeps its minimum value of $f_C\sim0$ from $B^{max}_{pro}\sim0$ to $B^{max}_{pro}\sim300$. As $B^{max}_{pro}$ increases from $B^{max}_{pro}\sim300$ to $B^{max}_{pro}\sim1200$, $f_C$ firstly increases and then decreases with the rise of $B^{max}_{pro}$. The maximum value of $f_C\sim 0.1$ is observed at the point of $B^{max}_{pro}\sim 600$. Within the range of $B^{max}_{pro}>1200$, $f_C$ keeps its stable value of $f_C\sim 0.06$. The critical point of $B^{maxc1}_{pro}\sim300$, above which the cooperators begin to occur, and the critical point of $B^{maxc2}_{pro}\sim1200$, above which the frequency of cooperators keeps stable, are observed. For a larger $B^{max}_I$, the critical point of $B^{maxc1}_{pro}$ has little change while the critical point of $B^{maxc2}_{pro}$ increases with the rise of $B^{max}_I$. The maximum value of $f_C$ increase with the rise of $B^{max}_{I}$.

\begin{figure}
\includegraphics[width=8cm]{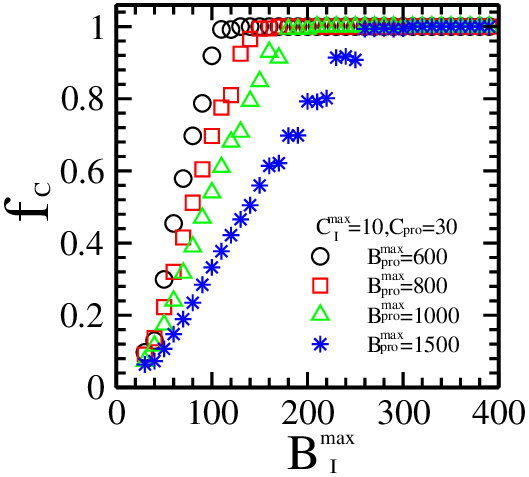}
\caption{\label{fig:epsart} the frequency of cooperators $f_c$ as a function of the upper limit of individual benefit $B^{max}_{I}$ for the upper limit of project benefit $B^{max}_{pro}$=600 (circles), 800 (squares), 1000 (triangles), 1500 (stars). Other parameters are: the needed project cost $C_{pro}$=30, the upper limit of individual contribution $C^{max}_{I}$=10, the minimum group size $n_{min}=2$, the maximum group size $n_{max}=50$, the total number of individuals $N$=1000. The data are obtained by averaging over 10 runs and $10^3$ time steps after $10^4$ relaxation time in each run.}
\end{figure}

The dependence of $f_c$ on $B^{max}_{I}$ for different $B^{max}_{pro}$ is presented in fig. 4. For a given $B^{max}_{pro}$=600, as $B^{max}_{I}$ increases from $B^{max}_{I}\sim 30$ to $B^{max}_{I}\sim 110$, $f_c$ increases from $f_c\sim 0.1$ to $f_c\sim 1$. As $B^{max}_{I}$ increases from $B^{max}_{I}\sim 110$ to infinity, $f_c$ keeps its maximum value of $f_c\sim 1$. The critical point of $B^{maxc}_{I}\sim 110$, above which $f_c$ does not change with the rise of $B^{max}_{I}$, is observed. An increase in $B^{max}_{pro}$ does not change the changing tendency of $f_c$ vs $B^{max}_{I}$ but leads to the rise of the critical point of $B^{maxc}_{I}$.

The results in fig. 3 and fig. 4 indicate that an optimal level of cooperation is determined by the coupling of the upper limit of individual benefit and the upper limit of project benefit. If an individual's benefit increases with the rise of the project benefit, an increase in the upper limit of project benefit promotes cooperation. If an individual's benefit has little change with the rise of the project benefit, an increase in the upper limit of project benefit does harm to cooperation. A larger $B^{max}_{I}$ and a comparatively small $B^{max}_{pro}$ lead to the prevalence of cooperation.

In the following, the evolution of the preferred group-sizes is explored and a functional relation between the frequency of cooperators and the distribution of the preferred group-sizes is found.

\begin{figure}
\includegraphics[width=10cm]{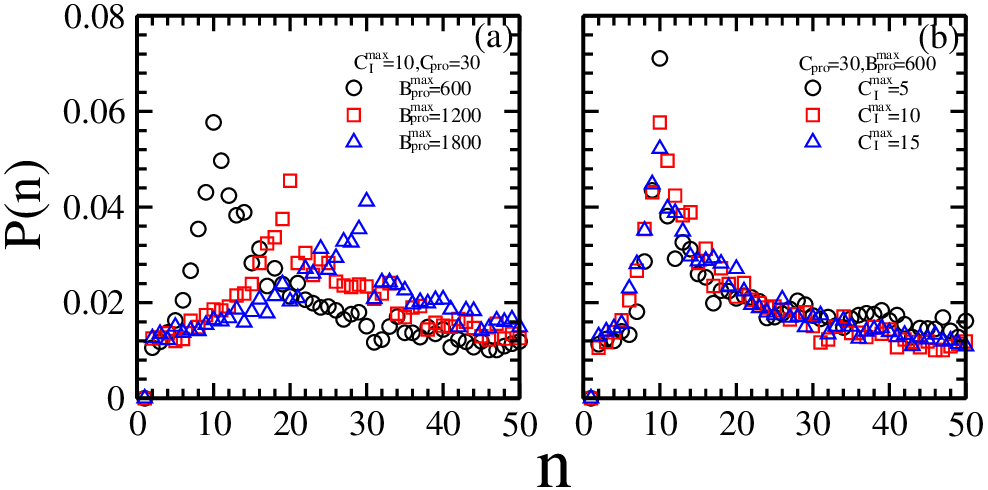}
\caption{\label{fig:epsart} the distribution of the preferred group-sizes (a) for the upper limit of project benefit $B^{max}_{pro}$=600 (circles), 1200 (squares), 1800 (triangles), and the needed project cost $C_{pro}=30$, the upper limit of individual contribution $C^{max}_{I}$=10, the upper limit of individual benefit $B^{max}_{I}=60$; (b) for the upper limit of individual contribution $C^{max}_{I}$=5 (circles), 10 (squares), 15 (triangles), and  the needed project cost $C_{pro}=30$, the upper limit of project benefit $B^{max}_{pro}$=600, the upper limit of individual benefit $B^{max}_{I}=60$. The data are obtained by averaging over 10 runs and $10^3$ time steps after $10^4$ relaxation time in each run.}
\end{figure}

In fig. 5 (a) and (b) we plot the distribution of the preferred group-sizes $P(n)$ for different values of the upper limit of project benefit $B^{max}_{pro}$ and different values of the upper limit of individual contribution $C^{max}_{I}$ respectively. Figure 5 (a) shows that, for $B^{max}_{pro}$=600 and $C^{max}_{I}$=10,$P(n)$ is like a Poisson distribution with the dominant group-size $n_{dom}\sim 10$. For $B^{max}_{pro}$=1200 and $C^{max}_{I}$=10,$P(n)$ is like a Poisson distribution with the dominant group-size $n_{dom}\sim 20$. For a larger $B^{max}_{pro}$, the distribution pattern of $P(n)$ changes little while the dominant group-size increases with the rise of $B^{max}_{pro}$. Figure 5 (b) shows that, for a given $B^{max}_{pro}$=600, an increase in the upper limit of individual contribution $C^{max}_{I}$ results in a broader distribution of the preferred group-sizes. The dominant group-size $n_{dom}\sim 10$ does not change with the rise of $C^{max}_{I}$.

\begin{figure}
\includegraphics[width=8cm]{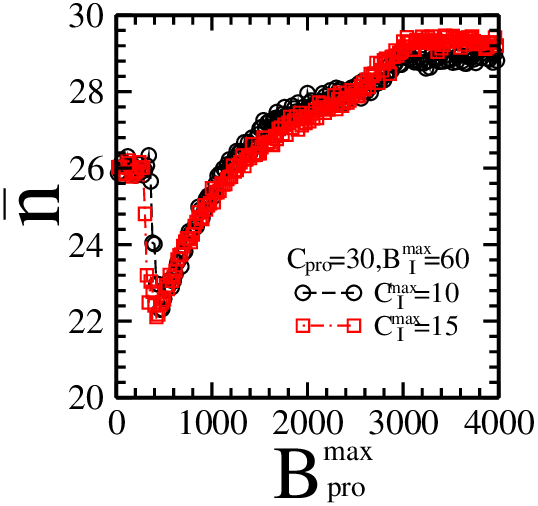}
\caption{\label{fig:epsart} the averaged value of the preferred group-sizes as a function of the upper limit of project benefit $B^{max}_{pro}$ for the upper limit of individual contribution $C^{max}_{I}$=10 (circles), 15 (squares). Other parameters are: the needed project cost $C_{pro}$=30, the upper limit of individual benefit $B^{max}_{I}$=60, the minimum group size $n_{min}=2$, the maximum group size $n_{max}=50$, the total number of individuals $N$=1000. The data are obtained by averaging over 10 runs and $10^3$ time steps after $10^4$ relaxation time in each run.}
\end{figure}

Figure 6 shows the dependence of the averaged value of the preferred group-sizes $\bar n$ on the upper limit of project benefit $B^{max}_{pro}$. It is observed that, within the range of $B^{max}_{pro}<300$, $\bar n$ keeps a constant value of $\bar n\sim 26$. As $B^{max}_{pro}$ increases from $B^{max}_{pro}=300$ to $B^{max}_{pro}=480$, $\bar n$ has a sharp drop from $\bar n\sim 26$ to $\bar n\sim 23$. As $B^{max}_{pro}$ increases from $B^{max}_{pro}\sim480$ to $B^{max}_{pro}\sim2000$, $\bar n$ increases continuously from $\bar n\sim 23$ to $\bar n\sim 28$. The critical point of $B^{maxc}_{pro}\sim300$ is observed. An increase in $C^{max}_{I}$ does not change the changing tendency of $\bar n$ vs $B^{max}_{pro}$ but leads to a decrease in the critical point of $B^{maxc}_{pro}$

The results in fig. 5 and fig. 6 indicate that both the distribution of the preferred group-size and the dominant group-size are determined by the coupling of the upper limit of individual contribution and the upper limit of project benefit. Comparing the results in fig. 1 with the results in fig. 6, we find that a larger value of the frequency of cooperators $f_C$ corresponds to a smaller value of the averaged value of the preferred group-size and a smaller value of the dominant group-size.

\begin{figure}
\includegraphics[width=8cm]{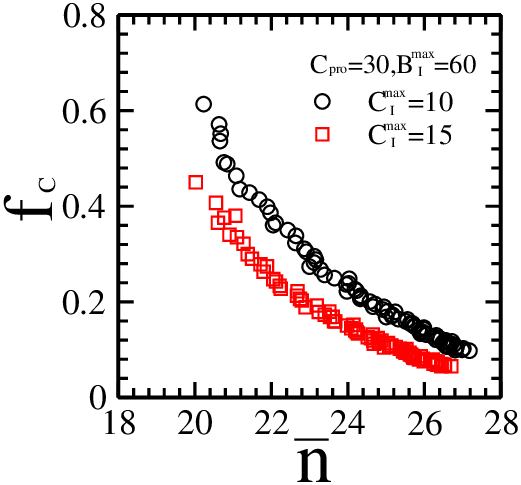}
\caption{\label{fig:epsart} the frequency of cooperators $f_C$ as a function of the averaged value of the preferred group-sizes for the upper limit of individual contribution $C^{max}_{I}$=10 (circles), 15 (squares). Other parameters are $C_{pro}$=30, $B^{max}_{I}$=10, the minimum group size $n_{min}=2$, the maximum group size $n_{max}=50$, the total number of individuals $N$=1000. The data are obtained by averaging over 10 runs and $10^3$ time steps after $10^4$ relaxation time in each run.}
\end{figure}

To find out the relationship between the frequency of cooperators $f_C$ and the distribution of the preferred group-sizes $P(n)$, the frequency of cooperators $f_C$ as a function of the averaged value of the preferred group-sizes $\bar n$ for different values of the upper limit of individual contribution $C^{max}_{I}$ are outlined in fig. 7. From fig. 7 we observe that, as the averaged value of the preferred group-sizes $\bar n$ increases from $\bar n\sim 21$ to $\bar n\sim 27$, the frequency of cooperators $f_C$ decreases from $f_C\sim0.62$ to $f_C\sim0.1$. As the upper limit of individual contribution $C^{max}_{I}$ becomes large, the downtrend of $f_C$ vs $\bar n$ becomes moderate. As we draw a line of best fit to the simulation data in fig. 7, a functional relation between $f_C$ and $\bar n$, that is $f_C\sim ae^{-b\bar n}$, $a\sim 114$ and $b\sim 0.259$ for $C^{max}_{I}$=10, $a\sim 157$ and $b\sim 0.291$ for $C^{max}_{I}$=15, is found.

In order to verify the robustness of the simulation results in the present model, we have checked the role of the population size and got the figures of $f_C$ vs $B^{max}_{pro}$ for different values of the population size $N$. The relation between $f_C$ and $B^{max}_{pro}$ is found to be robust against variations in the population size. We have also checked the role of the population size and the pre-given maximum group size $n_{max}$ in the distribution of the preferred group-size. It is found that the distribution of the preferred group-size is not affected by the population size. The pre-given maximum group size $n_{max}$ does not affect the dominant group size $n_{dom}$ but the distribution of the preferred group-size, which becomes broader as $n_{max}$ increases. In the present model, we pay attention to the role of individual ability and project benefit, therefore, we only consider the cases where $n_{max}$ is pre-given. The role of $n_{max}$ in the evolution of cooperation is another problem that deserves a deep research in the future. In addition to that, the frequency of cooperators $f_C$ is found to be closely related to the ratio of $\frac{C_{pro}}{C^{max}_I}$ and the ratio of $\frac{B^{max}_{pro}}{B^{max}_I}$. $f_C$ increases with the rise of $\frac{C_{pro}}{C^{max}_I}$ and decreases with the rise of $\frac{B^{max}_{pro}}{B^{max}_I}$.

In natural and human society, the common resource can be enjoyed by each member in the group, which may lead to the depletion of the common resource. Public cooperation is needed for the sustainability of the common resource. Depending upon the PGG and the NSG, the evolutionary mechanism of the occurrence and maintenance of cooperation among selfish individuals have been studied in the scenarios where the summed benefit of the common resource is limited and where the summed benefit of the common resource is unlimited respectively, but the coupling effect of the two scenarios on the evolution of cooperation is short of discussion. In the present model, by incorporating the parameters concerning an individual, i.e. the upper limit of individual contribution and the upper limit of individual benefit, and the parameters concerning a project, i.e. the needed project cost and the upper limit of project benefit, into the public goods game, the scenario in the PGG and that in the NSG are coupled. Therefore, the present model becomes interesting because it can be used to explore the evolutionary mechanism in a variety of situations which are consistent with the real world. The coupling effect of the maximum individual contribution and the needed project cost indicates that the individuals are more possible to cooperate as they face a much heavier task. The coupling effect of the upper limit of individual benefit and the upper limit of project benefit indicates that a higher level of cooperation can be sustained only if the project benefit is limited. The coevolution of the individual strategies and the group sizes indicates that a smaller group is beneficial for cooperation. The observations in the present work may be improved in the following two aspects. Firstly, the present model only considers the homogeneous situation where each agent has the same ability and all the projects that people done are the same. The evolution of cooperation in a heterogeneous environment deserves further study. Secondly, the feedback between the individual contribution and the project benefit has not been considered in the present model. How to incorporate the non-linear relation between the individual contribution and the project benefit into the present model is quite challenging.

\section{Theoretical analysis}
\label{sec:analysis}
\subsection{\label{subsec:levelA}Relationship between the dominant group size and the coupling of individual ability and project benefit.}
In the simulation results, it is found that the equilibrium frequency of cooperators is closely related to the averaged value of the preferred group-sizes, which is determined by the dominant group-size and the distribution of the preferred group-sizes. In the following, we give an analysis on the evolutionary mechanism of the dominant group-size in the present model.

Consider the case where the group-size $n$ is quite small, i.e. $n\le\frac{B_{pro}^{max}}{B^{max}_I}$. For $n_CC^{max}_I\ge C_{pro}$, the upper limit of project benefit is $B_{pro}^{max}$ and the payoffs of cooperators and defectors are $P_C=B^{max}_I-\frac{C_{pro}}{n_C}$ and $P_D=B^{max}_I$ respectively. For $n_CC^{max}_I< C_{pro}$, the project benefit is $B_{pro}=0$ and the payoffs of cooperators and defectors become $P_C=P_D=0$ accordingly. For a given $f_C$, the larger the group size $n$, the more possible there are $n_C(>\frac{C_{pro}}{C^{max}_I})$ cooperators in the group. Therefore, the preferred group-size $n$ tends to evolve to the maximum value of $n\sim\frac{B_{pro}^{max}}{B^{max}_I}$.

Consider the case where the group size $n$ is quite large, i.e. $n>\frac{B_{pro}^{max}}{B^{max}_I}$. For $n_CC^{max}_I\ge C_{pro}$, the upper limit of project benefit is $B_{pro}^{max}$ and the payoffs of cooperators and defectors are $P_C=\frac{B^{max}_{pro}}{n}-\frac{C_{pro}}{n_C}$ and $P_D=\frac{B^{max}_{pro}}{n}$ respectively. For $n_CC^{max}_I< C_{pro}$, the project benefit is $B_{pro}=0$ and the payoffs of cooperators and defectors become $P_C=P_D=0$ accordingly. An increase in the group size $n$ is more possible to lead to a decrease in $P_C$ and $P_D$. Therefore, the preferred group size tend to evolve to the minimum value of $n\sim\frac{B_{pro}^{max}}{B^{max}_I}$.

From the above analysis we find that the dominant group size is determined by the upper limit of project benefit and the upper limit of individual benefit. As the minimum value of the number of cooperators in the group is satisfied, i.e. $n_C\sim\frac{C_{pro}}{C^{max}_I}$, the dominant group size should be $n_{dom}\sim\frac{B_{pro}^{max}}{B^{max}_I}$, which is in accordance with the simulation data in fig. 5 (a).

\subsection{\label{subsec:levelB}Relationship between the equilibrium frequency of cooperators and the dominant group-size.}
The replicator dynamics can be applied to analyze the evolutionary behavior in n-player games\cite{zheng1}. From the above analysis we find that the dominant group-size in the present model is $n_{dom}=\frac{B_{pro}^{max}}{B^{max}_I}$ and the minimum value of the number of cooperators in the interacting group is $n_{C}^{min}=\frac{C_{pro}}{C^{max}_I}$.

Suppose that the system has evolved to the state where the preferred group-sizes of all the individuals are the same, i.e.$n=n_{dom}$,  and only the following three kinds of strategy combinations are left in the interacting group. Group one is composed of $n_{C}^{min}-1$ cooperators and $n_{dom}-n_{C}^{min}+1$ defectors. Group two is composed of $n_{C}^{min}$ cooperators and $n_{dom}-n_{C}^{min}$ defectors. Group three is composed of $n_{C}^{min}+1$ cooperators and $n_{dom}-n_{C}^{min}-1$ defectors. The payoffs of cooperators and defectors within these three groups are different from each other, that is, $P_C^{(1)}=P_D^{(1)}=0$ for group one, $P_C^{(2)}=B_{I}^{max}-\frac{C_{pro}}{n_{C}^{min}}$ and $P_D^{(2)}=B_{I}^{max}$ for group two, $P_C^{(3)}=B_{I}^{max}-\frac{C_{pro}}{n_{C}^{min}+1}$ and $P_D^{(3)}=B_{I}^{max}$ for group three.

Firstly, consider the case where only group two and group three coexist. For a cooperator in group two, because his payoff is less than the payoff of the cooperators in group three and the payoff of the defectors in group two and group three, that is $P_C^{(2)}<P_C^{(3)}<P_D^{(2)}=P_D^{(3)}$, he is more likely to become a defector or keep his original strategy. Therefore, group two is possible to become group one or have no change in the next time step. For a cooperator in group three, he is also more likely to become a defector or keep his original strategy. Therefore, group three is possible to become group two or have no change in the next time step. The above two evolutionary processes will lead to such a scenario where the number of group two keeps stable while the number of group three decreases.

Secondly, consider the case where only group one and group two coexist. For a cooperator or a defector in group one, because his payoff is less than the payoff of the cooperators or the defectors in group two, that is $P_C^{(1)}=P_D^{(1)}<P_C^{(2)}<P_D^{(2)}$, he is more likely to update his strategy and become a cooperator or defector as that in group two in the next time step. Therefore, group one is possible to become group two in the next time step. For a cooperator in group two, he is more likely to become a defector or keep his original strategy. Therefore, group two is possible to become group one in the next time step. The above two evolutionary processes will lead to such a scenario where the number of group two keeps stable while the number of group one decreases.

Therefore, as the system has evolved to the state where only group one, group two and group three are left, it is more likely that the system will reach an equilibrium in the next time step. The equilibrium frequency of cooperators should be $f_C\sim\frac{n^{min}_C}{n_{dom}}$, in which $n_C^{min}\sim\frac{C_{pro}}{C_I^{max}}$, $n_{dom}\sim\frac{B_{pro}^{max}}{B^{max}_I}$.

To verify the above analysis, here we only consider the simplest case where $n_{dom}=3$ and $n^{min}_C=2$. The averaged payoff of cooperators is

\begin{equation}
\label{eq.2}
\bar{P_C}=\frac{C_{N_c-1}^2(B^{max}_I-\frac{C_{pro}}{3})+C_{N_c-1}^1C_{N-N_c}^1(B^{max}_I-\frac{C_{pro}}{2})+{C_{N-N_c}^2\times0}}{C_{N-1}^2},
\end{equation}
and the averaged payoff of defectors is

\begin{equation}
\label{eq.2}
\bar{P_D}=\frac{C_{N_c}^2B^{max}_I+(C_{N_c}^1C_{N-N_c}^1+{C_{N-N_c}^2)\times0}}{C_{N-1}^2}.
\end{equation}
In the equilibrium state, $\bar{P_C}=\bar{P_D}$,

\begin{equation}
\label{eq.2}
C_{N_c-1}^2(B^{max}_I-\frac{C_{pro}}{3})+C_{N_c-1}^1C_{N-N_c}^1(B^{max}_I-\frac{C_{pro}}{2})={C_{N_c}^2B^{max}_I}.
\end{equation}

\begin{equation}
\label{eq.2}
\frac{1}{2}(N_c-1)(N_c-2)(B^{max}_I-\frac{C_{pro}}{3})+(N_c-1)(N-N_c)(B^{max}_I-\frac{C_{pro}}{2})=\frac{1}{2}N_c(N_c-1)B^{max}_I.
\end{equation}

\begin{equation}
\label{eq.2}
(N_c-2)(B^{max}_I-\frac{C_{pro}}{3})+2(N-N_c)(B^{max}_I-\frac{C_{pro}}{2})=N_cB^{max}_I.
\end{equation}

\begin{equation}
\label{eq.2}
(f_c-\frac{2}{N})(B^{max}_I-\frac{C_{pro}}{3})+2(1-f_c)(B^{max}_I-\frac{C_{pro}}{2})=f_cB^{max}_I,
\end{equation}

\begin{equation}
\label{eq.2}
(\frac{2C_{pro}}{3}-2B^{max}_I)f_c=C_{pro}-2B^{max}_I+\frac{2}{N}(B^{max}_I-\frac{C_{pro}}{3}),
\end{equation}

\begin{equation}
\label{eq.2}
f_c=\frac{2B^{max}_I-C_{pro}-\frac{2}{N}(B^{max}_I-\frac{C_{pro}}{3})}{2B^{max}_I-\frac{2C_{pro}}{3}},
\end{equation}

\begin{equation}
\label{eq.2}
f_c=\frac{1-\frac{C_{pro}}{2B^{max}_I}-\frac{1}{N}(1-\frac{C_{pro}}{3B^{max}_I})}{1-\frac{C_{pro}}{3B^{max}_I}}.
\end{equation}

On condition that $N\to\infty$, we get $f_c=\frac{1-\frac{C_{pro}}{2B^{max}_I}}{1-\frac{C_{pro}}{3B^{max}_I}}$. For a given $C_{pro}$ and $B_{pro}^{max}$, increasing  $B^{max}_I$ will lead to a decrease in $n_{dom}$ and an increase in $f_C$, which is in accordance with the theoretical analysis of $f_C\sim\frac{n^{min}_C}{n_{dom}}$.

\section{Summary}
\label{sec:summary}

In sum, we have put forward a generalized version of the public goods game with limited individual ability and maximum project rewards, focusing on revealing how the limitations of individual ability and project rewards affect the prevalence of cooperation. The lower the upper limit of an individual's contribution, the higher the frequency of cooperation. Such an impact is closely related to the upper limit of project benefit. The transition point of the upper limit of project benefit, at which the highest level of cooperation is obtained, is found.

The evolutionary dynamics of cooperation in the present model can be understood by taking into account the coevolution of the cooperator frequencies and the preferred group-sizes. As the upper limit of project benefit decreases, the frequency of cooperators increases while the averaged value of the preferred group-sizes decreases. A functional relation between the equilibrium frequency of cooperators and the dominant group sizes is found.

In future studies, it will be interesting to investigate how the heterogeneity of individual abilities influences the prevalence of cooperation. Some pioneering efforts have already been made along this path. Depending upon the evolutionary PD game, A. Szolnoki et al. have studied the effect of the heterogeneity of strategy adoption on the evolution of cooperation\cite{szolnoki20}. A significant increase in cooperation has been found. Depending upon the PD game, M. Perc et al. have studied the role of diversity in wealth and social status in the evolution of cooperation\cite{perc21}. They have found that the inhomogeneity in wealth and social status can effectively promote cooperation. Depending upon the evolutionary PD game, A. Szolnoki et al. have studied the role of the diversity of reproduction rate in the evolution of cooperation. They have found that the diversity in the reproduction capability is beneficial for the occurrence of cooperative behaviour\cite{szolnoki21}. In addition to that, the revealed mechanism of diverse sets of thresholds can be incorporated into other evolutionary game models and the evolutionary dynamics closer to the real world will be explored extensively.

\section*{Acknowledgments}
This work is the research fruits of National Natural Science Foundation of China (Grant Nos. 71371165, 71631005, 71273224, 71471161,71471031,71171036, 71202039, 61503109,10905023,71131007), Collegial Laboratory Project of Zhejiang Province (Grant No. YB201628),  Zhejiang Provincial Natural Science Foundation of China (Grant No. LY17G030024), Research Project of Generalized Virtual Economy(Grant No. GX2015 -1004(M)), Jiangxi Provincial Young Scientist Training Project (Grant No. 2013 3BCB23017). The major project of the National Social Science Foundation of China (Grant Nos. 12$\&$ZD067, and 14AZD089), the Program of Distinguished Professor in Liaoning Province (Grant No. [2013]204), the Project of Key Research Institute of Humanities and Social Sciences by Department of Education of Liaoning Province (Grant Nos. 15YJA790092, and 15YJC790041), and the Program of Discipline Support Plan in Dongbei University of Finance and Economics (Grant No. XKK-201401). The Fundamental Research Funds for the Central Universities.





\bibliographystyle{model1-num-names}



\end{document}